\documentclass
[showpacs,onecolumn,12pt,aps,pra,superscriptaddress,floatfix]
{revtex4-1}
\usepackage{graphicx}
\usepackage{dcolumn}
\usepackage{bm}
\usepackage{hyperref}
\usepackage{amsmath} 
\usepackage{color}
\usepackage{geometry}
\begin{document}

\title{Enhancement of coherent dipole coupling between two atoms via squeezing a cavity mode}

\author{Yan Wang}
\affiliation{Department of Physics, Harbin Institute of Technology, Harbin, 150001, China}

\author{Chuang Li}
\affiliation{Department of Physics, Harbin Institute of Technology, Harbin, 150001, China}

\author{Elijah M. Sampuli}
\affiliation{Department of Physics, Harbin Institute of Technology, Harbin, 150001, China}

\author{Jie Song} \email{E-mail: jsong@hit.edu.cn}
\affiliation{Department of Physics, Harbin Institute of Technology, Harbin, 150001, China}

\author{Yongyuan Jiang}
\affiliation{Department of Physics, Harbin Institute of Technology, Harbin, 150001, China}

\author{Yan Xia}
\affiliation{Department of Physics, Fuzhou University, Fuzhou, 350002, China}

\date{\today}

\begin{abstract}
We propose a theoretical method to enhance the coherent dipole coupling between two atoms in an optical cavity via parametrically squeezing the cavity mode. In the present scheme, conditions for coherent coupling are derived in detail and diverse dynamics of the system can be obtained by regulating system parameters. In the presence of environmental noise, an auxiliary squeezed field is employed to suppress, and even completely eliminate the additional noise induced by squeezing. In addition, we demonstrate that our scheme enables the effective suppression of atomic spontaneous emission. The results in our investigation could be used for diverse applications in quantum technologies.
\end{abstract}
\maketitle

\section{Introduction}
Transport of excitation via resonant interaction between
dipoles is fundamental to numerous disciplines, ranging from life sciences to quantum computing \cite{PhysRevLett.100.243201}. In quantum physics, the interaction between two atoms, one initially in the excited state and the other in the ground state, through virtual photon exchange is generally known as dipole-dipole interaction \cite{PhysRevA.89.043838}. It enables the atoms to resonantly exchange their energies by virtue of homenergic transition between energy levels. The availability of a strong long-range dipole-dipole interaction is an enabling resource for a wide range of studies and applications \cite{RevModPhys.82.2313,PhysRevA.96.042306,Quinten:98}. Typically, the interaction between atoms in highly excited Rydberg states attracts a great deal of research interests and has been elaborately studied both in theoretical schemes and in experimental realization \cite{0022-3700-14-21-003,PhysRevA.72.022347,PhysRevLett.93.153001,PhysRevLett.93.233001,PhysRevA.61.062309,labuhn2016tunable}. This strong interaction can give rise to the so-called dipole blockade mechanism \cite{PhysRevLett.100.113003,Comparat:10}, which has been demonstrated experimentally \cite{PhysRevLett.99.073002,gaetan2009observation,PhysRevLett.99.163601,PhysRevLett.105.193603,PhysRevLett.97.083003} and offers possibilities for generating entanglement of several atoms \cite{urban2009observation,PhysRevA.84.013831,PhysRevA.72.042302}, or implementing scalable quantum logic gates \cite{PhysRevLett.85.2208,PhysRevLett.87.037901} as well as other applications in quantum information processing \cite{PhysRevA.66.065403,PhysRevLett.99.260501}.

Cavity quantum electrodynamics (QED) studies the light-matter interactions at the quantum level in terms of single atoms coupled to quantized cavity field \cite{PhysRevA.97.043820,PhysRevLett.85.2392,PhysRevLett.83.5166}. Realization of the strong coupling regime (SCR) via high manipulation and control of quantum systems \cite{PhysRevLett.117.043601} is requisite for implementing quantum information tasks. However, strong strength of light-matter coupling requires resonators with high quality factors and small mode volumes simultaneously \cite{PhysRevLett.112.213602}, which remains extremely challenging to implement in experiments \cite{petersson2012circuit,PhysRevLett.105.140501}. Alternatively, flexible schemes for effectively enhancing atom-cavity coupling have been proposed. Recent studies have made it easy to reach the ultrastrong coupling (USC) or deep strong coupling (DSC) regime \cite{kockum2018ultrastrong,PhysRevLett.105.263603,sanchez2018resolution}. For instance, Leroux and co-workers proposed a method to enhance the qubit-cavity coupling via parametric driving of the cavity \cite{PhysRevLett.120.093602}. Qin \textit{et al.} exploited optical parametric amplification to enhance the atom-field coupling as well as the cooperativity of the system \cite{PhysRevLett.120.093601}. Upon parametrically squeezing the cavity mode, exponentially enhanced coupling strength has been achieved. This makes the squeezing-based scheme a powerful tool for enhancing coupling, while the dynamics of the system becomes complicated as a consequence. In-depth studies on the diverse dynamics are of importance and need to be promoted. 

For controlled quantum dynamics, it is prerequisite to achieve strong interactions between single pairs of atoms. Motivated by the recent advances mentioned above, here, we propose a scheme to enhance the coherent dipole coupling between two atoms confined in an optical cavity. The general idea is based on parametrically squeezing the cavity mode. We show that the resonance interaction between atoms can be achieved in the presence of squeezing, for the atoms coupled to cavity with both identical and different strengths. In both cases, we derive in detail the effective Hamiltonian to describe the dynamics and verify the parametric conditions for resonance interaction, as well as enhancement of coupling. For consideration of the environmental noise, the dynamics of the system will be destroyed due to the additional noise induced by squeezing. We show that the undesired noise can be suppressed and even completely eliminated by employing a squeezed field with proper parameters. In particular, we demonstrate that our scheme is capable of effectively suppressing the atomic spontaneous emission.

The remainder of the paper is structured as follows. In Sec.\,\ref{sec:model}, we describe the physical model of the system and give the Hamiltonian in the presence of squeezing. In Sec.\,\ref{sec:dynamics}, we show how squeezing the cavity mode enables the dramatical enhancement of the coherent coupling between atoms. In Sec.\,\ref{sec:noise}, the environmental noise is considered into the system. We present an approach to suppress the squeezing-induced noises by simply employing a squeezed field. Finally, we give a brief discussion on experimental implementations and summarize our conclusions in Sec.\,\ref{sec:conclusion}.

\section{The system and Hamiltonian}\label{sec:model}

\begin{figure}
	\centering
	\includegraphics[width=0.8\linewidth]{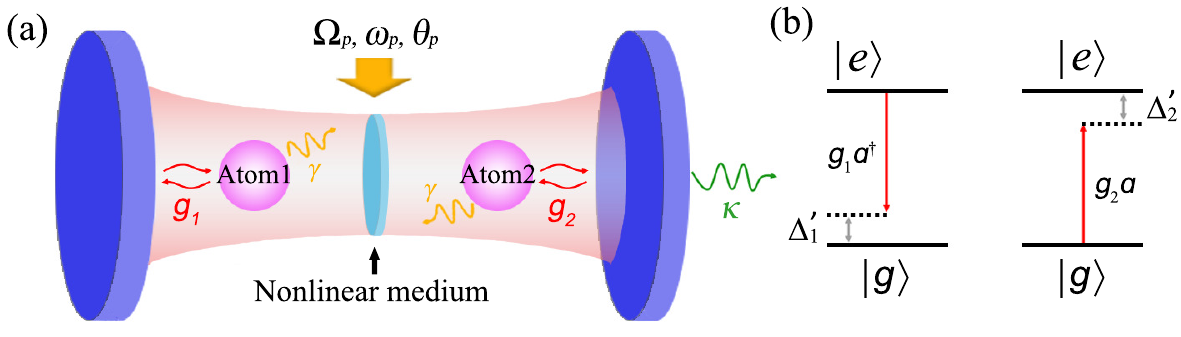}
	\caption{Schematics of the system. (a) Two atoms and a nonlinear medium are trapped in an optical cavity. The nonlinear medium is pumped at frequency $\omega_p$, amplitude $\Omega_p$ and phase $\theta_p$. The coupling strengths between two atoms and cavity mode are $g_1$ and $g_2$, respectively. The decay rates of atoms and cavity are $\gamma$ and $\kappa$, respectively. (b) Level scheme of two atoms interacting with cavity field with detunings $\Delta_1'$ and $\Delta_2'$ ($\Delta_i'=\omega_i-\omega_c$), respectively. }
	\label{Fig1}
\end{figure}

We consider a quantum system consisting of two identical two-level atoms and a nonlinear medium, which are confined into a single mode cavity (see Fig.\,\ref{Fig1}). The cavity mode can be squeezed while the nonlinear medium is pumped at frequency $\omega_p$, amplitude $\Omega_p$, and phase $\theta_p$. Working in a frame rotating at half the squeeze frequency $\omega_p/2$, the Hamiltonian of this system is given by (hereafter, $\hbar$=1)
\begin{equation}\label{Eq:original H}
H=\Delta_ca^\dag a+
\sum_{i=1,2}[\frac{\Delta_i}{2}\sigma_z^i+g_i(\sigma_+^ia+a^\dag\sigma_-^i)]+\frac{\Omega_p}{2}(e^{\mathrm{i}\theta_p}a^2+e^{-\mathrm{i}\theta_p}a^{\dag2}).
\end{equation}

Here, $a$($a^\dag$) is the annihilation(creation) operator of the cavity mode with frequency $\omega_c$. The two-level atom is described by the Pauli operator $\sigma_z^i=\vert e\rangle_i\langle e\vert-\vert g\rangle_i\langle g\vert$ ($i$=1,2 labels the atoms) and the transition operators $\sigma_+^i={\sigma_-^i}^\dag=\vert e\rangle_i\langle g\vert$, where $\vert e\rangle_i$ and $\vert g\rangle_i$ are the excited state and the ground state, respectively.  $g_i$ is the atom-cavity coupling strength. $\Delta_c=\omega_c-\omega_p/2$ and $\Delta_i=\omega_i-\omega_p/2$ are the cavity and atom detunings (with $\omega_i$ being the frequency of atom $i$). The Hamiltonian can be diagonalized by introducing the Bogoliubov squeezing transformation $a=a_s\cosh(r_p)-e^{-\mathrm{i}\theta_p}\sinh(r_p)a_s^\dag$ \cite{scully1999quantum}, where the squeezing parameter $r_p$ is controllable and defined via $r_p=(1/2)\arctan(\alpha)$ ($\alpha=\Omega_p/\Delta_c$). The resulting Hamiltonian can be expressed as follows
\begin{equation}\label{Eq2}
H'=\Delta_sa_s^\dag a_s+\sum_{i=1,2}[\frac{\Delta_i}{2}\sigma_z^i+g_i\cosh(r_p)(\sigma_+^ia_s+a_s^\dag\sigma_-^i)
-g_i\sinh(r_p)(e^{-\mathrm{i}\theta_p}\sigma_+^ia_s^\dag+e^{\mathrm{i}\theta_p}a_s\sigma_-^i)].
\end{equation}
The Hamiltonian (\ref{Eq2}) has the form of the usual Rabi Hamiltonian, where $\Delta_s=\Delta_c\sqrt{1-\alpha^2}$ denotes the squeezed cavity frequency. The effective coupling strengths in $H'$ show an enhancement of atom-cavity coupling, which has been verified for the single atom case in \cite{PhysRevLett.120.093601}. In contrast, our system focuses on the dipole transitions, i.e., state transfer between two atoms using the cavity mode as a quantum bus \cite{PhysRevA.97.032341}, and we intend to validate the possibility of enhancement of coherent dipole coupling between two atoms. To illuminate this, we then transform the Hamiltonian $H'$ into the interaction picture and obtain ($\theta_p$ is set to zero for simplicity)
\begin{equation}\label{Eq3}
\begin{split}
H_I=&g_1\cosh(r_p)a_s\sigma_+^1e^{\mathrm{i}\Delta_xt}+g_2\cosh(r_p)a_s\sigma_+^2e^{\mathrm{i}\Delta_yt}\\
&-g_1\sinh(r_p)a_s^\dag\sigma_+^1e^{\mathrm{i}\Delta_zt}-g_2\sinh(r_p)a_s^\dag\sigma_+^2e^{\mathrm{i}\Delta_wt}
+\mathrm{H.c.},
\end{split}
\end{equation}
where
\begin{equation}\label{Eq4}
\begin{split}
&\Delta_x=\Delta_1-\Delta_s,\ \Delta_y=\Delta_2-\Delta_s,\\
&\Delta_z=\Delta_1+\Delta_s,\ \Delta_w=\Delta_2+\Delta_s.
\end{split}
\end{equation}

We consider the states $\vert e_1g_2,n\rangle$ and $\vert g_1e_2,n\rangle$ with one excited atom and $n$ photons in the cavity. For large detunings with $\vert\Delta_x\vert,\vert\Delta_y\vert,\vert\Delta_z\vert,\vert\Delta_w\vert\gg \vert g_i\cosh(r_p)\vert,\vert g_i\sinh(r_p)\vert$, there is no energy exchange between atomic system and cavity. We can adiabatically eliminate the non-resonant states and obtain the effective Hamiltonian by using the perturbation theory. The coherent coupling between two atoms under different parametric conditions will be discussed separately in the following section.

\section{Dynamics and enhancement of coherent coupling}\label{sec:dynamics}
\subsection{Different atoms-cavity coupling strengths}
We first examine the dipole-dipole interaction or state transfer between the two atoms coupled to the cavity mode with unequal strengths, $g_1\neq g_2$. In addition, we assume that the detunings $\Delta_1$ and $\Delta_2$ are different; the cavity mode is initially in the vacuum state, which is only virtually excited due to the large detuning conditions. Thus the cavity mode will be always in the vacuum state. By adiabatically eliminating the cavity mode, we obtain the following effective atomic Hamiltonian \cite{doi:10.1139/p07-060}:
\begin{equation}\label{Eq5}
\begin{split}
H_{\mathrm{eff}}=&\frac{g_1^2\cosh^2(r_p)}{\Delta_x}\vert e\rangle_1\langle e\vert
+\frac{g_2^2\cosh^2(r_p)}{\Delta_y}\vert e\rangle_2\langle e\vert
-\frac{g_1^2\sinh^2(r_p)}{\Delta_z}\vert g\rangle_1\langle g\vert
-\frac{g_2^2\sinh^2(r_p)}{\Delta_w}\vert g\rangle_2\langle g\vert\\
&+\big[\frac{g_1g_2\cosh^2(r_p)}{2}(\frac{1}{\Delta_x}+\frac{1}{\Delta_y})\sigma_+^1\sigma_-^2e^{\mathrm{i}(\Delta_x-\Delta_y)t}
-\frac{g_1g_2\sinh^2(r_p)}{2}(\frac{1}{\Delta_z}+\frac{1}{\Delta_w})\sigma_+^1\sigma_-^2e^{\mathrm{i}(\Delta_z-\Delta_w)t}\\
&+\mathrm{H.c.}\big].
\end{split}
\end{equation}
The first four terms describe the photo-number-dependent Stark shifts, and the rest describe the dipole coupling between the two atoms. To realize efficient energy transition between states $\vert e_1g_2\rangle$ and $\vert g_1e_2\rangle$, the following condition should be satisfied
\begin{equation}\label{Eq6}
\begin{split}
\frac{g_1^2\cosh^2(r_p)}{\Delta_x}+\frac{g_1^2\sinh^2(r_p)}{\Delta_z}+\Delta_1=\frac{g_2^2\cosh^2(r_p)}{\Delta_y}+\frac{g_2^2\sinh^2(r_p)}{\Delta_w}+\Delta_2.
\end{split}
\end{equation}
	
\begin{figure}
	\centering
	\includegraphics[width=0.65\linewidth]{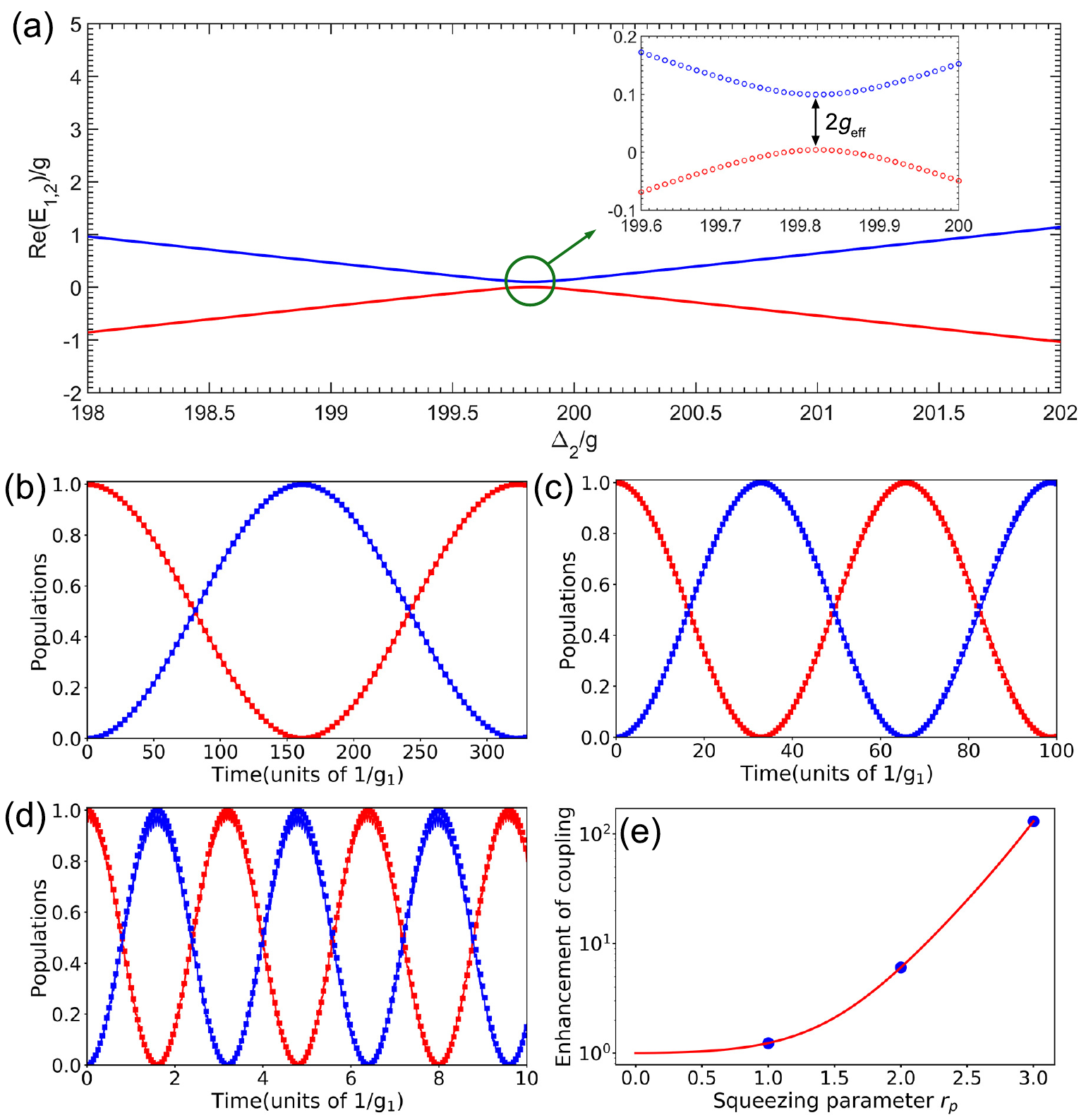}
	\caption{(a) Eigenenergies (real part of the eigenvalues $E_{1,2}$) as a function of detuning $\Delta_2$. Only the lowest two energy levels are shown. The inset shows a clear view of the avoided crossing phenomenon. The avoided crossing point corresponds to $\Delta_2=199.82g_1$, indicating the position where resonance occurs. The difference between the two levels is 0.096$g_1$ which is approximately equal to 2$g_{\mathrm{eff}}$. Here, the squeezing parameter $r_p$ is set as 2. (b)–(d) Oscillations between states $\vert e_1g_2\rangle$ (red curves) and $\vert g_1e_2\rangle$ (blue curves) obtained from the exact Hamiltonian (solid curves) and the effective Hamiltonian (square points) with different values of $r_p$: (b) $r_p=1$, (c) $r_p=2$, and (d) $r_p=3$. The shrink of oscillation period with increasing $r_p$ indicates the enhancement of coupling strength. (e) Enhancement of coupling as a function of $r_p$ ranging from 0 to 3. The three dots correspond to the cases in (b), (c), and (d), respectively. The common parameters in all figures are $g_2=1.5g_1$, $\Delta_1=200g_1$, and $\Delta_c=10g_1\cosh(r_p)/\sqrt{1-\alpha^2}$, and various values of $\Delta_2$ are obtained from Eq.\,(\ref{Eq6}). The initial state of the system is $\vert e_1g_2,0\rangle$.}
	\label{Fig2}
\end{figure}

The accuracy of this condition can be examined by calculating the eigenenergies of the states, which are obtained as the real part of the eigenvalues. In Fig.\,\ref{Fig2}(a), we plot the real part of the eigenvalues $E_{1,2}$ obtained from exact Hamiltonian (\ref{Eq2}) as a function of detuning $\Delta_2$. Here, only the lowest two energy levels are shown since others are well separated from the subspace. The eigenenergies of the two states avoid crossing at $\Delta_2=199.82g_1$ (where resonance occurs) and are split by $2g_{\mathrm{eff}}=0.096g_1$, as magnified in the inset in Fig.\,\ref{Fig2}(a). $g_{\mathrm{eff}}$ is expressed as $\frac{g_1g_2}{2}[\cosh^2(r_p)(\frac{1}{\Delta_x}+\frac{1}{\Delta_y})-\sinh^2(r_p)(\frac{1}{\Delta_z}+\frac{1}{\Delta_w})]$, which is obtained from effective Hamiltonian (\ref{Eq5}).
Simultaneously, by inserting the corresponding parameters into Eq.\,(\ref{Eq6}), the detuning $\Delta_2$ is calculated to be $199.82g_1$, which is in good accordance with the result shown in Fig.\,\ref{Fig2}(a). Thus, we confirm that Eq.\,(\ref{Eq6}) gives the parametric condition for effective resonance between the two atoms coupled to the cavity mode with different strengths.

In what follows, we plot the oscillations between the states $\vert e_1g_2\rangle$ and $\vert g_1e_2\rangle$ of the two atoms \cite{JOHANSSON20121760}, with squeezing parameter $r_p$ taken as several values, as depicted in Figs.\,\ref{Fig2}(b)–\ref{Fig2}(d). The solid curves correspond to the exact results obtained from the total Hamiltonian (\ref{Eq2}), whereas the square points denote the approximate results obtained from the effective Hamiltonian (\ref{Eq5}). Clearly, the approximate results agree well with the exact results when our proposed condition (\ref{Eq6}) is satisfied, which is a further evidence of the validity of the effective Hamiltonian $H_{\mathrm{eff}}$. Note that the effective coupling strength in Hamiltonian (\ref{Eq5}) depends on the squeezing parameter $r_p$. Therefore, we predict that the effective coupling will be largely enhanced with the increasing of $r_p$. By comparing the dynamics shown in Figs.\,\ref{Fig2}(b)–\ref{Fig2}(d), we can see that the period of oscillation decreases apparently with the increasing of $r_p$, corresponding to the enhancement of effective coupling strength. It is worth mentioning that the ideal Rabi-like oscillations exist for $r_p$ as large as 3 under the proposed parameters. With a higher value of $r_p$, the ideal oscillations will be destroyed because the large detuning conditions are no longer satisfied. This can be surmounted by properly increasing the atomic detuning $\Delta_1$, e.g., the ideal oscillations occur with $\Delta_1=600g_1$ for $r_p=4$. We further plot the enhancement of coupling as a function of $r_p$ ranging from 0 to 3, as shown in Fig.\,\ref{Fig2}(e). A strong enhancement of coupling exceeding $10^2$ can be achieved in the present parameter range.

\subsection{Identical atoms-cavity coupling strength}
We now discuss the dynamics of our system when the coupling strengths between the two atoms and the cavity are identical, namely, $g_1=g_2=g$. Under the parametric condition $\Delta_1=\Delta_2=\Delta$ (or $\Delta_x=\Delta_y$, $\Delta_z=\Delta_w$), the effective Hamiltonian (\ref{Eq5}) can be simplified to
\begin{equation}\label{Eq7}
\begin{split}
H_{\mathrm{eff}}=&\frac{g^2\cosh^2(r_p)}{\Delta_x}\sum_{i=1,2}\vert e\rangle_i\langle e\vert-\frac{g^2\sinh^2(r_p)}{\Delta_z}\sum_{i=1,2}\vert g\rangle_i\langle g\vert\\&+\big[\big(\frac{g^2\cosh^2(r_p)}{\Delta_x}-\frac{g^2\sinh^2(r_p)}{\Delta_z}\big)\sigma_+^1\sigma_-^2+\mathrm{H.c.}\big].
\end{split}
\end{equation}

\begin{figure}
	\centering
	\includegraphics[width=0.8\linewidth]{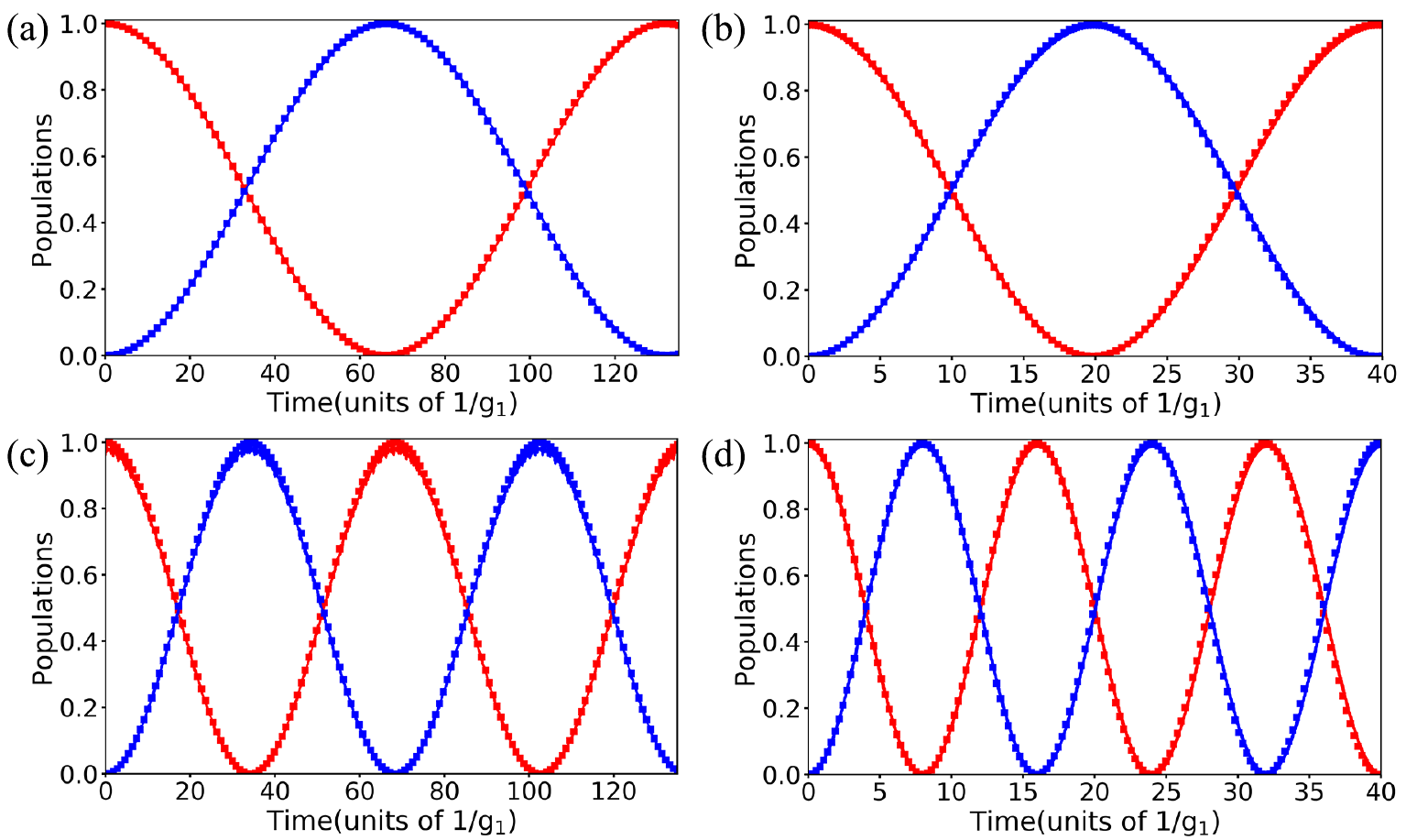}
	\caption{Oscillations between states $\vert e_1g_2\rangle$ (red curves) and $\vert g_1e_2\rangle$ (blue curves) obtained from the exact Hamiltonian (solid curves) and the effective Hamiltonian (square points). The squeezing parameters $r_p$ in (a),(b) and (c),(d) are set as 1 and 2, respectively. Other parameters in (a),(c) are $\Delta=50g$, $\Delta_s=0.05\Delta$; in (b),(d), $\delta=20g\cosh(r_p)$, and the values of $\Delta_s$ and $\Delta$ fulfill  $\Delta_s-\Delta=\delta$ and $\Delta_s+\Delta=20\delta$. The initial state of the system in all figures is $\vert e_1g_2,0\rangle$.}
	\label{Fig3}
\end{figure}

For different values of detuning $\Delta_s$, the efficient energy transfer between the two atoms can be realized. Here, we mainly consider two conditions associated with $\Delta_s$: (1) $\Delta_s$ is far less than $\Delta$; (2) the difference between $\Delta_s$ and $\Delta$ is equal to a certain value $\delta$ that is much larger than $\{g\cosh(r_p),g\sinh(r_p)\}$, and the sum of $\Delta_s$ and $\Delta$ is much larger than $\delta$. The oscillations between the states $\vert e_1g_2\rangle$ and $\vert g_1e_2\rangle$ of the two atoms under conditions (1) and (2) are plotted in Figs.\,\ref{Fig3}(a) and \ref{Fig3}(b), respectively. The solid curves and square points are obtained by using the total Hamiltonian (\ref{Eq2}) and effective Hamiltonian (\ref{Eq7}), respectively. As clearly shown in the figures, the dynamics of the system exhibit ideal Rabi-like oscillations for $\Delta_s$ in different parametric regimes. To verify whether the effective coupling strengths in Hamiltonian (\ref{Eq7}) (under the above-mentioned two conditions) can be enhanced by increasing the squeezing parameter $r_p$, we plot the corresponding dynamics with $r_p=2$, as depicted in Figs.\,\ref{Fig3}(c) and \ref{Fig3}(d). By separately comparing the results shown in the left or right panel of Fig.\,\ref{Fig3}, it can be directly found that the period of oscillation decreases almost a half as $r_p$ increases from 1 to 2. These results reveal that we indeed can realize the enhancement of dipole coupling by simply increasing the squeezing parameter. Overall, the effect of enhancement is valid in broad parametric space where the atoms-cavity coupling strengths and detunings take various values. Furthermore, we note that Eq.\,(\ref{Eq6}) is a versatile condition because it is also valid for $g_1=g_2$ and $\Delta_1=\Delta_2$.

Next, we consider a special condition $\Delta=0$. In this case, the Hamiltonian (\ref{Eq3}) transforms to 
\begin{equation}
H_I=g\cosh(r_p)a_s\sum_{i=1,2}\sigma_+^ie^{-\mathrm{i}\Delta_st}
-g\sinh(r_p)a_s^\dag\sum_{i=1,2}\sigma_+^ie^{\mathrm{i}\Delta_st}+\mathrm{H.c.}.
\end{equation}

\begin{figure}[h]
	\centering
	\includegraphics[width=0.8\linewidth]{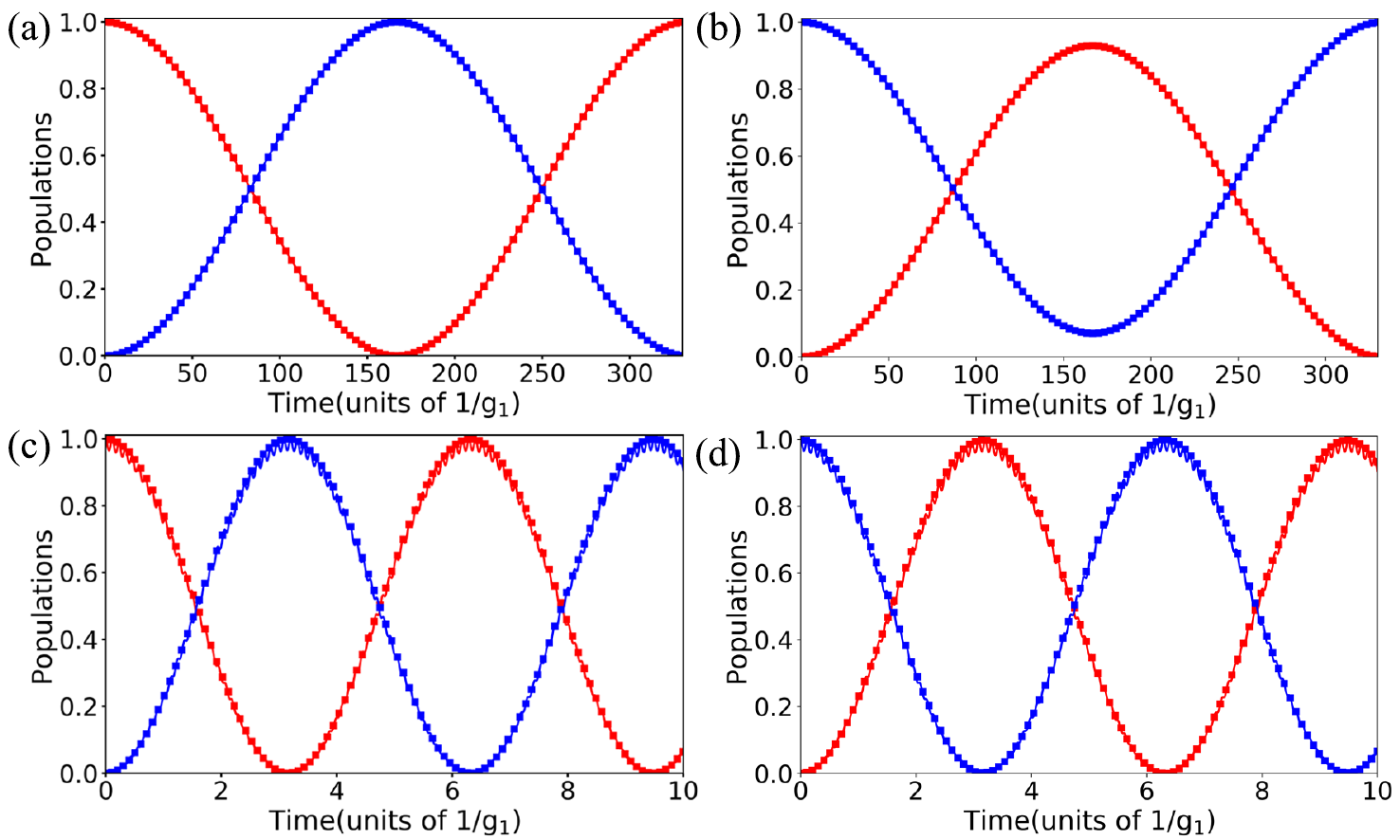}
	\caption{Oscillations between states (a),(c) $\vert e_1g_2\rangle$ and $\vert g_1e_2\rangle$ (b),(d) $\vert e_1e_2\rangle$ and $\vert g_1g_2\rangle$ obtained from the exact Hamiltonian (solid curves) and the effective Hamiltonian (square points). The squeezing parameters $r_p$ in (a),(b) and (c),(d) are set as 1 and 2, respectively. Other common parameters in (a–d) are $\Delta=0$ and $\Delta_c=1500g$. The initial state of the system in (a),(c) and (b),(d) are $\vert e_1g_2,0\rangle$ and $\vert g_1g_2,0\rangle$, respectively.}
	\label{Fig4}
\end{figure}

By using the perturbation theory and adiabatically eliminating the cavity mode, the effective atomic Hamiltonian can be written as
\begin{equation}\label{Eq9}
\begin{split}
H_{\mathrm{eff}}=&-\frac{g^2}{\Delta_s}\big[\cosh^2(r_p)\sum_{i=1,2}\vert e\rangle_i\langle e\vert+\sinh^2(r_p)\sum_{i=1,2}\vert g\rangle_i\langle g\vert\big]\\
&+\frac{g^2}{\Delta_s}\big[\cosh(2r_p)\sigma_+^1\sigma_-^2-\sinh(2r_p)\sigma_+^1\sigma_+^2+\mathrm{H.c.}\big].
\end{split}
\end{equation}
As expected, we can see the terms describing the transitions between $\vert e_1g_2\rangle$ and $\vert g_1e_2\rangle$ in the Hamiltonian (\ref{Eq9}). More interestingly, the specific condition $\Delta=0$ results in new terms $\sigma_+^1\sigma_+^2$ and $\sigma_-^1\sigma_-^2$ appearing in the Hamiltonian, corresponding to the transition path $\vert e_1e_2\rangle$ $\leftrightarrow$ $\vert g_1g_2\rangle$. The dynamics of these two transition paths with different squeezing parameters are shown in Fig.\,\ref{Fig4}. For the path $\vert e_1g_2\rangle$ $\leftrightarrow$ $\vert g_1e_2\rangle$, coherent population oscillation occurs for $r_p=1$ [see Fig.\,\ref{Fig4}(a)] and the period of it shrinks markedly when $r_p$ increases to 2 [see Fig.\,\ref{Fig4}(c)]. However, for the path $\vert e_1e_2\rangle$ $\leftrightarrow$ $\vert g_1g_2\rangle$, an inadequate oscillation between $\vert e_1e_2\rangle$ and $\vert g_1g_2\rangle$ is observed for $r_p=1$ [see Fig.\,\ref{Fig4}(b)], arising from the asymmetrical energy shifts associated with $r_p$ in the Hamiltonian (\ref{Eq9}). With the increasing of $r_p$, the effective coupling can be enhanced to much larger than the difference between the energy shifts. This compensates the inadequate oscillation and therefore the ideal Rabi-like oscillation reappears in the dynamics, as shown in Fig.\,\ref{Fig4}(d).

\begin{figure}
	\centering
	\includegraphics[width=0.6\linewidth]{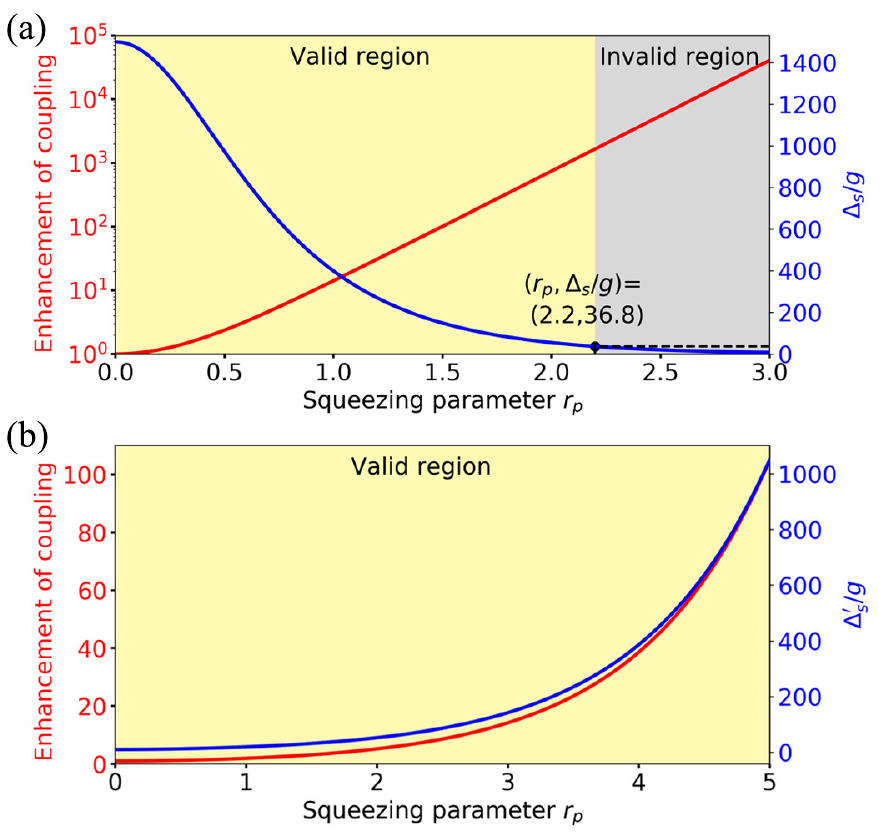}
	\caption{Validity of the enhancement of effective coupling. (a) The detuning $\Delta_s/g$ (blue curve) and the enhancement of coupling (red curve, expressed as $\cosh(2r_p)/\sqrt{1-\alpha^2}$) as functions of $r_p$, when $\Delta_c$ is set as $1500g$. The yellow and gray shaded areas represent the valid ($r_p<2.2$, where ideal oscillations can be obtained) and the invalid region ($r_p\ge2.2$, where ideal oscillations will be destroyed) of the enhancement of coupling, respectively. The black dot corresponds to $r_p=2.2$ and $\Delta_s/g=36.8$. (b) The detuning $\Delta_{s}^{'}/g$ (blue curve, set as $10\sqrt{\cosh(2r_p)}$) and the enhancement of coupling (red curve, expressed as $\sqrt{\cosh(2r_p)}$) as functions of $r_p$. Based on the adjustment of detuning, the valid region of the enhancement of coupling is enlarged to $r_p=5$ (and even more).}
	\label{Fig5}
\end{figure}

The above discussions are valid under the condition $\Delta_c=1500g$. When, however, the squeezing parameter $r_p$ further increases to over 2, the ideal Rabi-like oscillations will be destroyed. This is reasonable because the large detuning condition $\Delta_s\gg\{g\cosh(r_p),g\sinh(r_p)\}$ cannot be satisfied perfectly with the increasing of $r_p$. To give more insights into this, we take the transition $\vert e_1g_2\rangle$ $\leftrightarrow$ $\vert g_1e_2\rangle$ as an example, and plot the detuning $\Delta_s/g$ and the enhancement of coupling as functions of $r_p$ in Fig.\,\ref{Fig5}(a). Clearly, $\Delta_s/g$ drops off rapidly with the increasing of $r_p$ and down to near zero when $r_p$ increases to 3. By checking the numerical results of the dynamical evolutions, it can be directly seen that the ideal oscillations will be destroyed when $r_p$ reaches and exceeds 2.2 (not shown here). Accordingly for $r_p=2.2$, $\Delta_s/g$ is calculated to be 36.8, which is close to tenfold of $\{\cosh(r_p),\sinh(r_p)\}$. This indicates that the enhancement of coupling is valid in the region $r_p<2.2$ while invalid in the region $r_p\ge2.2$, as labeled in Fig.\,\ref{Fig5}(a). To enlarge the valid region of the enhancement of coupling, a possible strategy is to change the detuning $\Delta_s$ to $10g\sqrt{\cosh(2r_p)}$ ($\Delta_{s}^{'}$). In Fig.\,\ref{Fig5}(b), we plot the detuning $\Delta_{s}^{'}/g$ and the corresponding enhancement of coupling as functions of $r_p$ ranging from 0 to 5. Upon the adjustment of detuning, the large detuning condition can be always satisfied due to the incremental increase of $\Delta_{s}^{'}/g$ with increases in $r_p$. Therefore, the valid region of the enhancement of coupling, where the ideal Rabi-like oscillations occur (verified by the numerical simulations of the dynamics), is greatly enlarged as labeled in Fig.\,\ref{Fig5}(b).

\section{Consideration of squeezing-induced noise}\label{sec:noise}

\begin{figure*}
	\centering
	\includegraphics[width=0.9\linewidth]{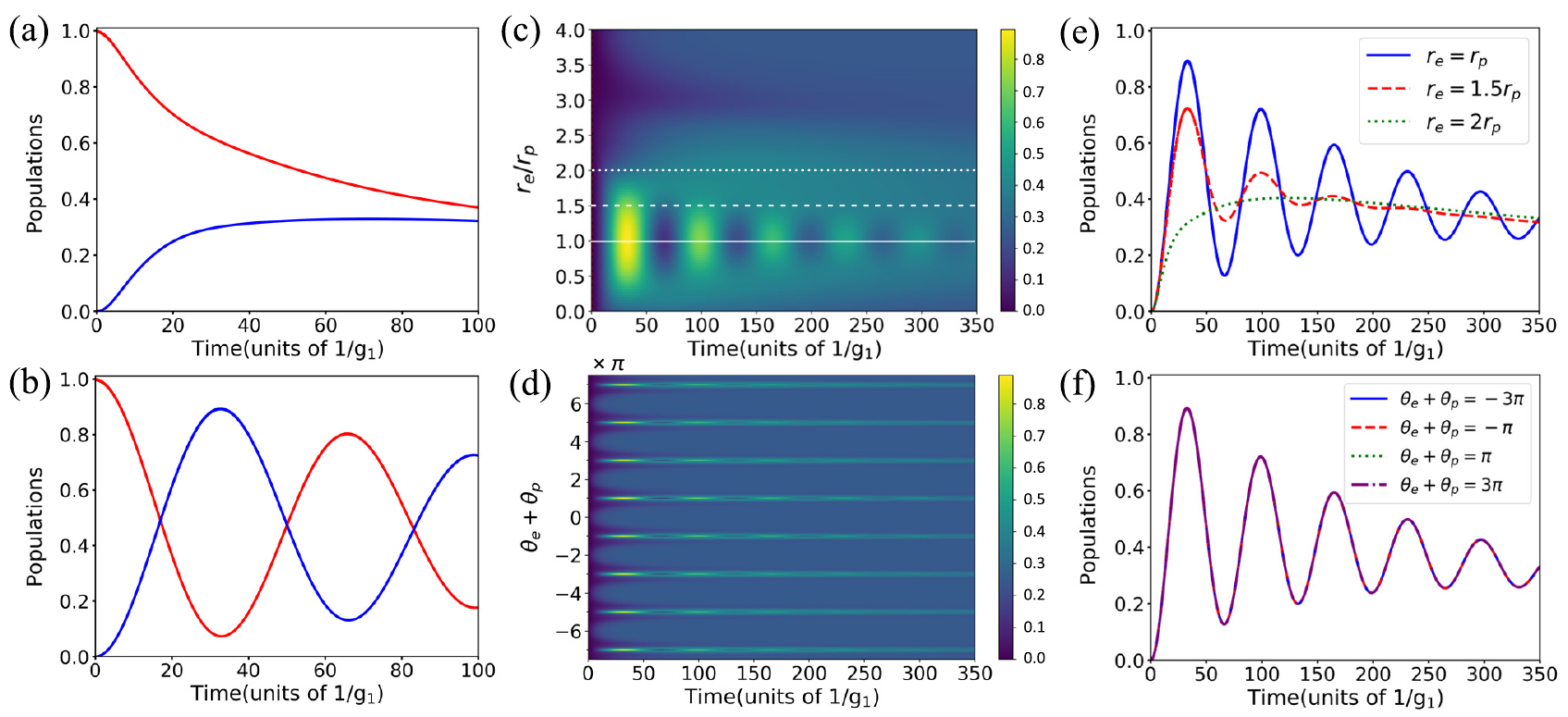}
	\caption{Dynamics of the system under consideration of the environmental noise. (a),(b) Oscillations between states $\vert e_1g_2\rangle$ and $\vert g_1e_2\rangle$ before and after introducing the squeezed field, which are obtained by numerically solving the master equation (\ref{Eq:master equation without squeezed field}) and (\ref{Eq:master equation with squeezed field}), respectively. (c),(d) Oscillations of the state $\vert g_1e_2\rangle$ as functions of $r_e/r_p$ and $\theta_e+\theta_p$, respectively. The three horizontal lines in (c) denote the $r_e/r_p$ values of 1, 1.5, and 2 (from bottom to top), and the corresponding oscillations are clearly plotted in (e). (f) Oscillations of the state $\vert g_1e_2\rangle$ for $\theta_e+\theta_p=-3\pi, -\pi, \pi$, and $3\pi$. The parameters in all figures are $g_2=1.5g_1$, $r_p=2$, $\Delta_1=200g_1$, $\Delta_2=199.822g_1$ [obtained from Eq.\,(\ref{Eq6})], $\Delta_c=10g_1\cosh(r_p)/\sqrt{1-\alpha^2}$, $\kappa=\gamma=10^{-3}g_1$, $\bar{n}_{\mathrm{th}}=5$, $\theta_p=0$; in (b), (c), and (e), $\theta_e=\pi$; in (b), (d), and (f), $r_e=2$.}
	\label{Fig6}
\end{figure*}

We have demonstrated the enhancement of coherent dipole coupling between the two atoms by  parametrically squeezing the cavity mode. However, the introduction of squeezing will lead to additional noise into the system. To demonstrate explicitly the squeezing-induced noise, it is straightforward to derive the master equation in the presence of squeezing. Here, we assume that the cavity mode is coupled to a thermal reservoir. When the cavity mode is squeezed with squeezing parameter $r_p$ and reference phase $\theta_p$, the system master equation in terms of $a_s$ is derived as (see Appendix for details)
\begin{equation}\label{Eq:master equation without squeezed field}
\begin{split}
\dot{\rho}(t)=&i[\rho(t),H(t)]-\frac{1}{2}\Big\{\sum_{x=1,2}\mathcal{L}(L_{x})\rho(t)
+[(2N+1)\bar{n}_{\mathrm{th}}+N+1]\mathcal{L}(L_{as})\rho(t)\\
&+[(2N+1)\bar{n}_{\mathrm{th}}+N]\mathcal{L}(L_{as}^\dag)\rho(t)
-[M(2\bar{n}_{\mathrm{th}}+1)]\mathcal{L'}(L_{as}^\dag)\rho(t)
-[M^{\ast}(2\bar{n}_{\mathrm{th}}+1)]\mathcal{L'}(L_{as})\rho(t)
\Big\},
\end{split}
\end{equation}
where $H(t)$ is the Hamiltonian given by Eq.\,(\ref{Eq:original H}); $N$ and $M$ are derived as
\begin{subequations}
\begin{equation}
N=\sinh^2(r_p),
\end{equation}
\begin{equation}
M=\cosh(r_p)\sinh(r_p)e^{\mathrm{-i}\theta_p}.
\end{equation}
\end{subequations}
These two parameters describe the effective thermal noise and two-photon correlation of the squeezed cavity mode \cite{PhysRevLett.114.093602,breuer2002theory}. 
The introduction of these noises will destroy the regular system dynamics, i.e., suppressing the amplitude of oscillations. To illustrate this, we take the oscillations between states $\vert e_1g_2\rangle$ and $\vert g_1e_2\rangle$ [described by Hamiltonian (\ref{Eq5})] as an example. In Fig.\,\ref{Fig6}(a), we numerically solve the master equation (\ref{Eq:master equation without squeezed field}) and plot the oscillations between states $\vert e_1g_2\rangle$ and $\vert g_1e_2\rangle$, which are immensely suppressed due to the additional noise induced by cavity mode squeezing.

It is demonstrated in Ref.\,\cite{PhysRevLett.120.093601} that the squeezing-induced noise can be suppressed by introducing an auxiliary squeezed-vacuum field to drive the cavity. Motivated by this, we explore the noise-resisted scheme in our system by virtue of the squeezed field. The squeezed field with the squeezing parameter $r_e$ and reference phase $\theta_e$, which has a much larger linewidth than the cavity mode, can be regarded as a squeezed reservoir. In this situation, we can assume that the cavity mode is coupled to a squeezed thermal reservoir. By choosing appropriate matching conditions, e.g., $r_e=r_p$ and $\theta_e+\theta_p=\pi$, the squeezing-induced noise can be completely eliminated (i.e., $N_s=\bar{n}_{\mathrm{th}}$ and $M_s=0$, see Appendix for details) and the dynamic of system is therefore governed by the master equation in the standard Lindblad form
\begin{equation}\label{Eq:master equation with squeezed field}
\dot{\rho}(t)=i[\rho(t),H'(t)]-\frac{1}{2}\Big\{\sum_{x=1,2}\mathcal{L}(L_{x})\rho(t)+(\bar{n}_{\mathrm{th}}+1)\mathcal{L}(L_{as})\rho(t)+\bar{n}_{\mathrm{th}}\mathcal{L}(L_{as}^\dag)\rho(t)
\Big\}.
\end{equation}
From the point of view of system-reservoir coupling, squeezing the cavity mode induces an increase in system-reservoir coupling strength. Employing an auxiliary squeezed field can, in principle, be equivalent to offset the increase in system-reservoir coupling via reservoir manipulation (squeezing), as proper matching conditions are satisfied.
In Fig.\,\ref{Fig6}(b), we plot the oscillations between states $\vert e_1g_2\rangle$ and $\vert g_1e_2\rangle$ by numerically solving the master equation (\ref{Eq:master equation with squeezed field}). As expected, the populations of states $\vert e_1g_2\rangle$ and $\vert g_1e_2\rangle$ exhibit rather strong oscillations.

For further insight into the parametric ranges for suppressing the squeezing-induced noise, we plot the dynamics of the same model as dependent on $r_e/r_p$ ranging from 0 to 4 in Fig.\,\ref{Fig6}(c). It reveals that the oscillations occur in a broad region except for the above-mentioned condition $r_e=r_p$. Specifically, the amplitude of oscillation during the whole evolution process reaches a peak for $r_e=r_p$ [marked with a solid line in Fig.\,\ref{Fig6}(c)], which gradually drops off with increasing or decreasing the value of $r_e/r_p$. A clearly view of the oscillations for several values of $r_e/r_p$ (1, 1.5, and 2) can be seen in Fig.\,\ref{Fig6}(e). Likewise, we also plot the dynamics as a function of $\theta_e+\theta_p$ in Fig.\,\ref{Fig6}(d). The oscillations periodically occur when $\theta_e+\theta_p=\pm n\pi\ (n=1, 3, 5 \cdots)$, and exhibit almost identical evolution behavior. To verify this, in Fig.\,\ref{Fig6}(f) we plot the oscillations of state $\vert g_1e_2\rangle$ for $\theta_e+\theta_p=-3\pi, -\pi, \pi$ and $3\pi$. It is clearly to see the exactly overlap of the oscillations.

\begin{figure}
	\centering
	\includegraphics[width=0.8\linewidth]{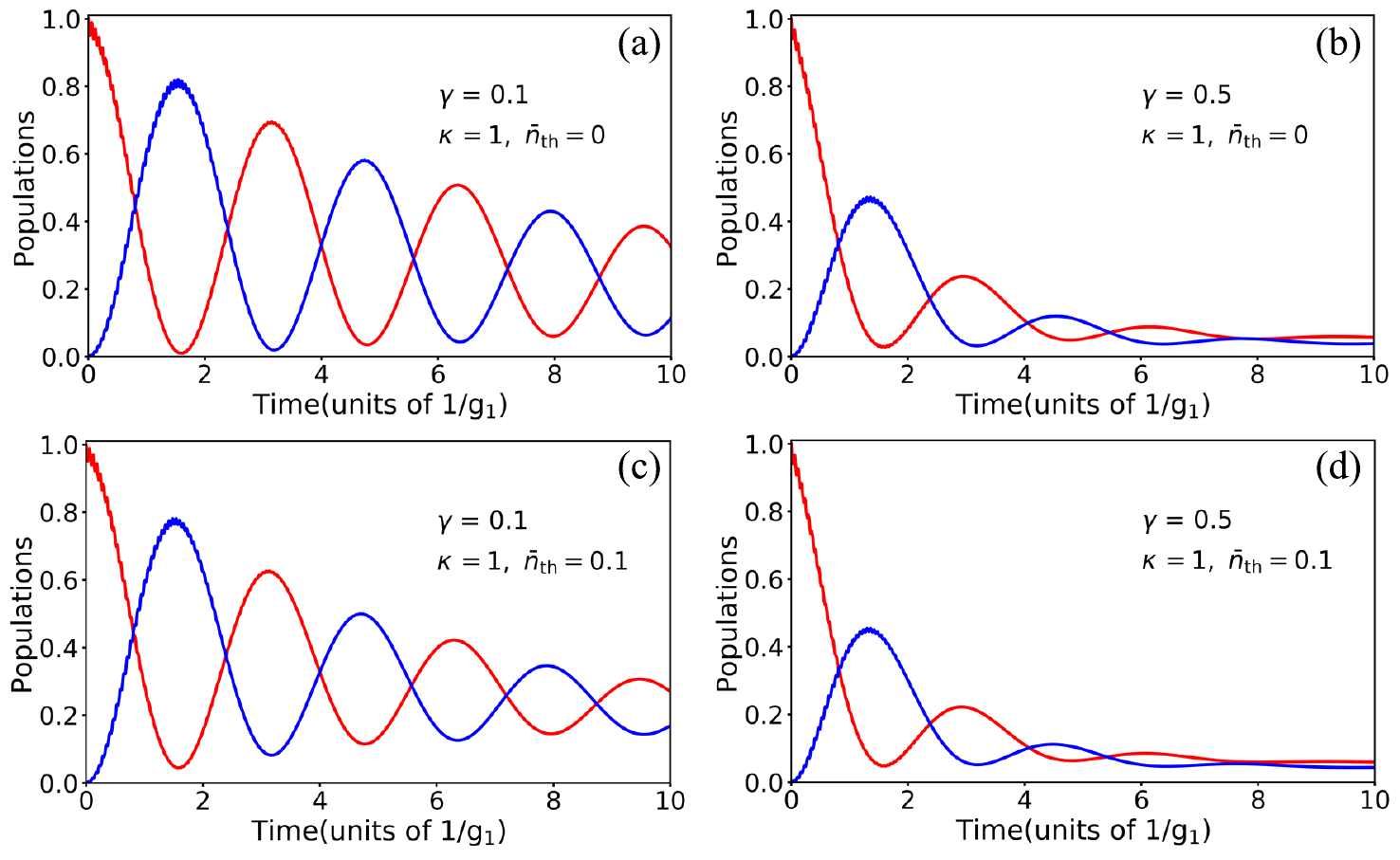}
	\caption{Oscillations between states $\vert e_1g_2\rangle$ (red curves) and $\vert g_1e_2\rangle$ (blue curves) obtained by solving the master equation (\ref{Eq:master equation with squeezed field}) at different values of decay rates and thermal photons (as labeled in figures). The parameters are $r_p=r_e=3$, $\theta_e=\pi$, and other common parameters are the same as Fig.\,\ref{Fig2}(d).}
	\label{Fig7}
\end{figure}

Another prominent feature of our method is the effective suppression of atomic spontaneous emission. As mentioned above, the coherent coupling between atoms can be greatly enhanced by squeezing the cavity mode, with the additional noise in the cavity being suppressed by applying the squeezed field. With the enhancement of coupling, the atomic dissipation cannot be affected and thus unchanged, which is equivalent to an effective suppression of atomic dissipation. This allows the observation of oscillations even for a large atomic decay rate $\gamma$, for instance, $\gamma/g_1=0.1$ [Fig.\,\ref{Fig7}(a)]. When we further increase $\gamma/g_1$ to 0.5, it can be seen that the oscillations also occur for several periods [Fig.\,\ref{Fig7}(b)]. Note that the cavity decay rate ($\kappa/g_1$) and thermal photon number ($\bar{n}_{\mathrm{th}}$) in Figs.\,\ref{Fig7}(a) and \ref{Fig7}(b) have been set as 1 and 0, respectively. The capacity of our scheme for resisting strong cavity dissipation is attributed to the smaller population of photons arising from the adiabatical elimination of the non-resonant intermediate state $\vert g_1g_2,1\rangle$. For comparison, we also plot the oscillations in the case of $\bar{n}_{\mathrm{th}}=0.1$, as shown in Figs.\,\ref{Fig7}(c) and \ref{Fig7}(d). Clearly, injection of the thermal photons results in a slight depopulation of states.

\section{Discussion and conclusion}\label{sec:conclusion}
We give a brief discussion regarding the experimental implementation. 
The promising pathway for realizing our proposal is based on the platform of cavity QED.
The configuration of the two-level atom can be realized in alkali-metal atoms, e.g., cesium \cite{PhysRevLett.97.083003,PhysRevLett.93.233603} and rubidium \cite{PhysRevLett.112.043601,PhysRevLett.94.033002}.
The single-mode cavity can be implemented typically using a high-finesse Fabry-Perot resonator \cite{PhysRevLett.118.133604,PhysRevLett.110.223003}.
An atom-cavity coupling strength ($g$) of tens of MHz is generally available.
The cavity mode squeezing can be generated via the process of optical parametric amplification, e.g., pumping a second-order nonlinear medium [periodically-poled potassium titanyl phosphate (PPKTP) crystal \cite{PhysRevLett.117.110801,schnabel2017squeezed}].
For producing the squeezed field with high bandwidth up to GHz, similar methods with generating cavity-field squeezing have been demonstrated in \cite{Ast:13,Serikawa:16} by using the PPKTP crystal.
The squeezing parameter and reference phase can be controlled via adjusting the amplitude and phase of the pumped laser.
On the basis of the above setups, we assume the following parameters: $g_1/2\pi=5$ MHz,  $g_2/2\pi=7.5$ MHz,  $r_p=r_e=2$, $\theta_p=0$, $\theta_e=\pi$, $\kappa/2\pi=500$ kHz, $\gamma/2\pi=5$ kHz, and $\bar{n}_{\mathrm{th}}=0.1$. 
We then choose the detunings $\Delta_1=200g_1$, $\Delta_2=199.822g_1$, and $\Delta_c=10g_1\cosh(r_p)/\sqrt{1-\alpha^2}$ to satisfy the resonant condition between the atoms. 
The resulting period of oscillation is about 2.1 $\mu$s and the population of state $\vert g_1e_2\rangle$ for the first period reaches 0.9, which we have verified numerically. 
In addition, our scheme is also promising for application in solid-state systems, in particular, superconducting quantum circuits that combine superconducting qubits with microwave-frequency cavity \cite{PhysRevLett.101.253602,PhysRevLett.119.023602}.

In conclusion, we have demonstrated that parametric squeezing of the cavity mode enables a strong enhancement of coherent dipole coupling between two atoms. The resonance between atoms occurs for atoms coupled to the cavity with both identical and different strengths, manifesting as the observation of coherent population oscillations in the dynamics. In both cases, we have derived in detail the effective Hamiltonian to describe the dynamics of various systems and verified the parametric conditions for effective state transfer, as well as the enhancement of coupling. It is demonstrated that the shrink of the period of oscillations with increasing the squeezing parameters can act as an indication of the enhancement of coupling. We hereby anticipate that by modulating the squeezing parameter, the enhancement can be achieved in broad parametric regimes where the atoms-cavity coupling strength and detuning take various values. For consideration of the environmental noise, the dynamics of the system will be destroyed due to the additional noise induced by squeezing, which can be completely eliminated by employing a squeezed field. The parametric conditions for resisting noise have been investigated and demonstrated here in detail. In addition, we also shown that our scheme can effectively suppress atomic spontaneous emission and cavity decay. The oscillations occur for several periods even for strong atomic and cavity dissipations. Our method for enhancing dipole-dipole interaction between atoms can be applicable to a wide range of physical systems, and will find various applications in quantum information processing.

\section{ACKNOWLEDGMENTS}
This work was supported by National Natural Science Foundation of China (NSFC) (11675046), Program for Innovation Research of Science in Harbin Institute of Technology (A201412), and Postdoctoral Scientific Research Developmental Fund of Heilongjiang Province (LBH-Q15060).

\appendix
\section*{APPENDIX: Effective master equations}
\setcounter{equation}{0}
\renewcommand{\theequation}{A\arabic{equation}}
Here, we first assume that the cavity mode is coupled to a thermal reservoir. Considering several relaxation processes (including the decay of atoms and cavity field) affecting the system, the dynamics of the system can be described by the quantum master equation in the standard Lindblad form \cite{scully1999quantum}
\begin{equation} \label{Eq:standard Lindblad form}
\dot{\rho}(t)=i[\rho(t),H(t)]-\frac{1}{2}\Big\{\sum_{x=1,2}\mathcal{L}(L_{x})\rho(t)+(\bar{n}_{\mathrm{th}}+1)\mathcal{L}(L_{a})\rho(t)+\bar{n}_{\mathrm{th}}\mathcal{L}(L_{a}^\dag)\rho(t)
\Big\},
\end{equation}
where $\rho(t)$ is the density operator of the system, $H(t)$ is the Hamiltonian given by Eq.\,(\ref{Eq:original H}), $L_a=\sqrt{\kappa}a$ is the Lindblad operator describing the cavity decay with rate $\kappa$, $L_{1,2}=\sqrt{\gamma}\vert g\rangle_{1,2}\langle e\vert$ are the Lindblad operators describing the atomic spontaneous emissions with identical rate $\gamma$, $\bar{n}_{\mathrm{th}}=(e^{\hbar\omega/k_BT}-1)^{-1}$ is the mean number of thermal photons in the cavity at temperature $T$, $\mathcal{L}(O)$ and $\mathcal{L}'(O)$ are defined by
\begin{subequations}
\begin{equation}
\mathcal{L}(O)\rho(t)=O^\dag O\rho(t)-2O\rho(t) O^\dag +\rho(t) O^\dag O,
\end{equation}
\begin{equation}
\mathcal{L}'(O)\rho(t)=O O\rho(t)-2O\rho(t) O+\rho(t)OO.
\end{equation}
\end{subequations}

When the cavity mode is squeezed with a squeezing parameter $r_p$ and a reference phase $\theta_p$, the master equation can be rewritten by simply performing Bogoliubov transformation $a=a_s\cosh(r_p)-e^{-\mathrm{i}\theta_p}\sinh(r_p)a_s^\dag$, given by
\begin{equation}\label{Eq in Appen:master equation without squeezed field}
\begin{split}
\dot{\rho}(t)=&i[\rho(t),H'(t)]-\frac{1}{2}\Big\{\sum_{x=1,2}\mathcal{L}(L_{x})\rho(t)+[(2N+1)\bar{n}_{\mathrm{th}}+N+1]\mathcal{L}(L_{as})\rho(t)\\
&+[(2N+1)\bar{n}_{\mathrm{th}}+N]\mathcal{L}(L_{as}^\dag)\rho(t)-[M(2\bar{n}_{\mathrm{th}}+1)]\mathcal{L'}(L_{as}^\dag)\rho(t)-[M^{\ast}(2\bar{n}_{\mathrm{th}}+1)]\mathcal{L'}(L_{as})\rho(t)
\Big\},
\end{split}
\end{equation}
where $H'(t)$ is given by Eq.\,(\ref{Eq2}), $\mathcal{L}_{as}=\sqrt{\kappa}a_s$ is the Lindblad operator describing the decay of the squeezed-cavity mode, $N$ and $M$ describe the effective thermal noise and two-photon correlation \cite{PhysRevLett.114.093602,breuer2002theory}, respectively, given by
\begin{subequations}
	\begin{equation}
	N=\sinh^2(r_p),
	\end{equation}
	\begin{equation}
	M=\cosh(r_p)\sinh(r_p)e^{\mathrm{-i}\theta_p}.
	\end{equation}
\end{subequations}

In the following, we demonstrate the squeezing-induced noise, i.e., thermal noise and two-photon correlation, can be suppressed by employing a squeezed field to drive the cavity. In view of a high-bandwidth squeezed field (up to GHz), which has been realized in experiments \cite{Ast:13,Serikawa:16}, it can be regarded as a squeezed reservoir due to a relatively small linewidth of typical optical cavity mode. Therefore, we assume that the cavity mode is coupled to a squeezed thermal reservoir with a squeezing parameter $r_e$ and phase $\theta_e$. The dynamics of system can be described by the following master equation \cite{doi:10.1080/09500349808230898}

\begin{equation}\label{EqA1}
\begin{split}
\dot{\rho}(t)=&i[\rho(t),H(t)]-\frac{1}{2}\Big\{\sum_{x=1,2}\mathcal{L}(L_{x})\rho(t)+(N'+1)\mathcal{L}(L_a)\rho(t)+N'\mathcal{L}(L_a^\dag)\rho(t)\\
&-M'\mathcal{L}'(L_a^\dag)\rho(t)-M'^\ast\mathcal{L}'(L_a)\rho(t)\Big\},
\end{split}
\end{equation}
where $N'$ and $M'$ are parameters that describe the squeezed thermal reservoir and are given by
\begin{subequations}
\begin{equation}
N'=\bar{n}_{\mathrm{th}}\cosh(2r_e)+\sinh^2(r_e),
\end{equation}
\begin{equation}
M'=(2\bar{n}_{\mathrm{th}}+1)\cosh(r_e)\sinh(r_e)e^{\mathrm{i}\theta_e}.
\end{equation}
\end{subequations}

Following the same method as before, the master equation after squeezing the cavity mode is re-expressed as
\begin{equation}\label{EqA5}
\begin{split}
\dot{\rho}(t)=&i[\rho(t),H'(t)]-\frac{1}{2}\Big\{\sum_{x=1,2}\mathcal{L}(L_{x})\rho(t)+(N_s+1)\mathcal{L}(L_{as})\rho(t)+N_s\mathcal{L}(L_{as}^\dag)\rho(t)\\
&-M_s\mathcal{L}'(L_{as}^\dag)\rho(t)-M_s^\ast\mathcal{L}'(L_{as})\rho(t)\Big\},
\end{split}
\end{equation}
where $N_s$ and $M_s$ are given by 
\begin{subequations}
\begin{equation}
\begin{split}
N_s=&[\bar{n}_{\mathrm{th}}\cosh(2r_e)+\sinh^2(r_e)]\cosh(2r_p)+\sinh^2(r_p)\\&+(\bar{n}_{\mathrm{th}}+\frac{1}{2})\sinh(2r_e)\sinh(2r_p)\cos(\theta_e+\theta_p),
\end{split}
\end{equation}
\begin{equation}
\begin{split}
M_s=&-\exp(-\mathrm{i}\theta_p)(2\bar{n}_{\mathrm{th}}+1)\Big\{\frac{1}{2}\sinh(2r_p)\cosh(2r_e)\\&+\frac{1}{2}\sinh(2r_e)\{\exp[\mathrm{i}(\theta_e+\theta_p)]\cosh^2(r_p)\\&+\exp[-\mathrm{i}(\theta_e+\theta_p)]\sinh^2(r_p)\}\Big\}.
\end{split}
\end{equation}
\end{subequations}

These two terms indicate the undesired noise in the cavity induced by squeezing, which can be removed by choosing appropriate condition. For instance, when choosing $r_e=r_p$ and $\theta_e+\theta_p=\pi$, the $N_s$ and $M_s$ can be reduced to $\bar{n}_{\mathrm{th}}$ and 0, respectively. In this case, the master equation (\ref{EqA5}) is simplified to the standard Lindblad form
\begin{equation}
\dot{\rho}(t)=i[\rho(t),H'(t)]-\frac{1}{2}\Big\{\sum_{x=1,2}\mathcal{L}(L_{x})\rho(t)+(\bar{n}_{\mathrm{th}}+1)\mathcal{L}(L_{as})\rho(t)+\bar{n}_{\mathrm{th}}\mathcal{L}(L_{as}^\dag)\rho(t)
\Big\}.
\end{equation}

\bibliography{reference.bib}

\end{document}